\documentclass[prd,aps,showpacs,epsbox,groupedaddress,eqsecnum,notitlepage,nofootinbib]{revtex4-1}
\usepackage[dvipdf]{graphicx}
\usepackage{epstopdf}
\usepackage{amsmath}
\usepackage{slashed}
\usepackage{amssymb}
\usepackage{enumerate, color}
\usepackage{subfigure}
\usepackage{tabularx}
\usepackage{epstopdf}
\usepackage{lipsum}
\usepackage{color}
\usepackage[utf8]{inputenc}

\newcommand{\be}{\begin{equation}}
\newcommand{\ee}{\end{equation}}
\newcommand{\ben}{\begin{eqnarray}}
\newcommand{\een}{\end{eqnarray}}
\newcommand{\bes}{\begin{subequations}}
\newcommand{\ees}{\end{subequations}}

\begin{document}

\title{Graphene wormhole trapped by external magnetic field}
\author{G. Q. Garcia$^{1}$\footnote{gqgarcia99@gmail.com }}
\author{P. J. Porf\'irio$^{2}$\footnote{pporfirio89@gmail.com }}
\author{D. C. Moreira$^{3}$\footnote{moreira.dancesar@gmail.com }}
\author{C. Furtado$^{4}$\footnote{furtado@fisica.ufpb.br}} 
\affiliation{$^1$Unidade Acadêmica de F\'isica, Universidade Federal da Campina Grande, 58109-970 Campina Grande, PB, Brazil}
\affiliation{$^2$Departament of Physics and Astronomy, University of Pennsylvania, Philadelphia, PA 19104, USA}
\affiliation{$^3$Faculdade Uninassau Petrolina, 56308-210, Petrolina, PE, Brazil}
\affiliation{$^4$Departamento de F\'isica, Universidade Federal da Para\'iba, 58051-970, Jo\~ao Pessoa, PB, Brazil}

\begin{abstract}
In this work we study the behavior of massless fermions in a graphene wormhole and in the presence of an external magnetic field. The graphene wormhole is made from two sheets of graphene  which play the roles of asymptotically flat spaces connected through a carbon nanotube with a zig-zag boundary. We solve the massless Dirac equation  within this geometry, analyze  the corresponding  wave function,  and show that the energy spectra of these solutions exhibit  behavior  similar to Landau levels.
\end{abstract}


\maketitle

\section{Introduction}

In recent years graphene has played a very important role in physics due to its incredible mechanical and electronic properties  allowing for many technological applications  \cite{geim2010rise, geim2009graphene}. Graphene  has  a carbon-honeycomb structure and constitutes the first really  two-dimensional material developed due to its unusual  conical  band structure, where the electrons close the Fermi points have a conical band structure ,
 being described by the massless Dirac equation (or Weyl equation) in (2 + 1)-dimensions \cite{bena2009remarks, neto2009electronic}. Therefore, a quasiparticle on graphene has a relativistic behavior, although it is not really a relativistic particle. Another interesting  effect is the presence of topological defects in the graphene lattice, since it is virtually impossible to prepare a real solid crystal without them. By means of "cut and glue" processes, Volterra studied these defects and their respective topological numbers, analyzing  changes occurring in the graphene lattice in the presence of pentagon rings, heptagon rings or pentagon-heptagon pairs \cite{volterra1907equilibre,moraes2000condensed,puntigam1997volterra}. Rings of pentagons or  heptagons are related to disclinations, and pentagon-heptagon pairs are related to dislocations \cite{carpio2008dislocations, vozmediano2010gauge}. A geometric description of defects in 3D solids based on Riemann-Cartan geometry was presented in the works of Katanaev and Volovich \cite{katanaev1992theory, katanaev2005geometric}. They  proved  an equivalence between a three-dimensional gravitational theory with torsion and the theory of defects in solids, associating curvature with disclinations and torsion with dislocations. Therefore, in the continuum limit, topological defects in crystals can be described in terms of Riemann-Cartan geometry.  It is well known  that the presence of disclinations in graphene modifies its electronic properties and that these topological defects can be interpreted as a source of a non-abelian gauge field, while smooth deformations of graphene sheets produce a gauge field similar to an electromagnetic  one \cite{vozmediano2010gauge, gonzalez1993electronic, garcia2017geometric}. In addition, another direct consequence of the presence of topological defects in graphene is that the gauge fields affects the quantum particle response to magnetic fields, causing its energy spectrum to present degenerate levels  analogous  to Landau levels \cite{lammert2004graphene, bueno2012landau}. For all these aspects, we can say that graphene is a  real laboratory for field theory with several possibilities to study the behavior of particles in $(2+1)$-dimensional systems presenting topological defects.

Many gravitational analogue models  based on using graphene in curved background geometries were constructed by exploring the geometric effects on the quantum dynamics of graphene quasiparticles. An interesting work on this subject was presented by Gonz\'alez {\it et al.} \cite{gonzalez1993electronic}, in which the quantum spectrum of a fullerene molecule is described through a graphene sheet  folded in a polyhedra. In this case, the graphene lattice  includes pentagonal rings, acting as a gauge field, and presents spherical symmetry. In 2007, Cortijo and Vozmediano proposed a model in which graphene is slightly curved and has pentagon-heptagon pairs and Stone-Wales defects in their structure, based on an analogy with cosmic strings \cite{cortijo2007electronic, cortijo2007effects}. In \cite{vozmediano2010gauge}, Vozmediano  discussed the physical properties of graphene, showing its connection with Quantum Field Theory (QFT) and Gauge Theory.  An interesting link between topological defects in graphene and cosmic strings in General Relativity (GR) also was presented. The  arising of geometric phases in graphitic nanocons was explored in \cite{furtado2008geometric}, and recently some of the authors  of \cite{furtado2008geometric} have analyzed the effects of rotation on fullerenes molecules in background geometries given by Gödel-type universes \cite{garcia2017geometric, garcia2017fermions, garcia2018weyl}. Some studies on dislocations and their relation to torsion theory have also been performed. In particular, the formation and stability of dislocations was studied by Carpio {\it et al.} \cite{carpio2008dislocations}, and Mesaros demonstrated the emergence of a non-Abelian Berry phase due to the existence of dislocations in the graphene structure \cite{mesaros2009berry}.

An important class of geometries  arising in gravitational models are wormholes. They have a long history, and their  properties  deserve some discussion. The first wormhole solution was put forward by Einstein and Rosen in the seminar paper \cite{einstein1935particle}. Such a solution is sometimes  called Einstein-Rosen bridge. These solutions are obtained by making a suitable coordinate change on the Schwarzschild geometry. This coordinate change covers two asymptotically identical flat spaces connected by a "bridge". In particular, Einstein-Rosen bridge presents event horizons, and as a result, observers are not allowed to travel along the bridge from a particular flat space to the other. On the other hand, it is possible to construct traversable wormholes,  so that  the two-way travel of observers  is permitted. For that, it is necessary to implement further conditions in order to prevent the emergence of the event horizons. As shown by Morris and Thorne in \cite{morris1988wormholes}, once these conditions are fulfilled, the wormhole is traversable. However, in GR, traversable wormholes are supported by matter sources violating the energy conditions \cite{morris1988wormholes}.   In this way, a wormhole is defined geometrically as a bridge between two manifolds or between two distant parts of the same manifold.  The wormholes can be Lorentzian (traversable or not), Euclidean, etc. It is well known in General Relativity that traversable Lorentzian wormholes can be obtained as solutions of Einstein  equations that require  so called exotic matter, i.e., matter violating the Null Energy Condition (NEC) \cite{morris1988wormholes,visser1995lorentzian}, which is a part of the Weak Energy Condition (WEC) whose physical meaning is that the
energy density is nonnegative in any reference frame. In addition to aforementioned properties, it is worth  to call attention  to other remarkable issues, such  as the construction of a time-machine from wormholes \cite{lobo2017wormholes} and inherently their causality violation problems \cite{visser1995lorentzian}, duality between wormholes and entangled states in quantum gravity \cite{maldacena2013cool} and analogue models (condensed matter systems used to mimic wormholes),  see \cite{visser2002analogue} and references therein.  Along these lines, analogue models have earned  increased attention since the beginning of GR, mainly in the last decades, typically in condensed matter systems. Such models,  in some sense, are attempts to describe the gravitational physics in terms of other physical systems. As pointed out in \cite{visser2002analogue}, the analogue models mimic kinematic properties of GR  but not the  dynamical ones. In particular, we shall mainly concentrate on a particular analogue model called graphene wormhole \cite{ gonzalez2010graphene,gonzalez2009propagating} whose properties will be given in details later. Furthermore, we focus on kinematic aspects of the graphene wormhole. In other words, we are concerned about recovering the geometrical properties of gravitational wormholes. Roughly speaking,  the graphene wormhole should be comprised of two graphene sheets (play a role of asymptotically flat spaces) connect by a carbon nanotube (playing a role of the bridge) in order to mimic the gravitational wormhole geometry. On the other hand, dynamic properties are not mimicked. For example, it is well known that gravitational wormholes violate the energy conditions in GR, and an issue that could be raised would be: should graphene wormholes  satisfy such conditions? The answer is obvious: no! Since the energy conditions are inherent to GR, the graphene wormhole does not mimic them. In \cite{park2013thermal}, the stability of geometrical structure of graphene was studied and  stable graphene wormhole structures were obtained by simulation procedure in the context of molecular dynamics. 

In the last decade several papers have dealt with a number of models of wormholes in carbon structures and other condensed matter systems.  One of the first models in this context was used to examine nanoporous carbon structures in \cite{margine2007theory}. Gonzalez and Herrero, by connecting two graphene sheets through a graphene nanotube, constructed an interesting model where the continuum limit of the graphene lattice is associated with a wormhole metric \cite{gonzalez2010graphene,gonzalez2009propagating}. Pincak and Smotlacha \citep{pincak2013analogies} have studied the electronic properties  of a graphene wormhole in a perturbed nanocylinder \citep{pincak2013analogies}, and  the electronic structure of a wormhole geometry in solid state models was investigated by Atanasov and Saxena \cite{atanasov2011electronic}. Fernandes {\it et al.} have analysed the wormhole geometry for electron waves in graphene \cite{fernandes2014wormhole}. Sepehri {\it et al.} proposed a model to describe the evolution of the free electron current density in graphene \cite{sepehri2017current} and Capozziello {\it et al.} presented a model which simulates moving electrons in graphene structures and analyzed its properties in Chern-Simons wormholes \cite{capozziello2018constructing}. One of the authors recently investigated the parallel transport of spinors in graphene wormhole and the Aharonov-Bohm analogue effect in this system \cite{carvalho2013holonomy}. Dandoloff and colleagues analyzed a geometry-induced potential in a two-dimensional model of wormhole geometry \cite{dandoloff2010geometry}, and another analogue wormhole model in porous carbon/magnetic composite particle was presented in \cite{fang2016wormhole}. 

In this work we focus on the study of wormhole-like geometries constructed from carbon nanostructures and on the exact solutions of the Dirac equation in $(2 + 1)$-dimensions in this setup \footnote{During the submission process we were informed about the  work \cite{Rojjanason2018icy}, which also gets Landau levels for fermions in a wormhole geometry. However, we must point out that  our study is different.}. In particular, we deal with the graphene wormhole model described by Gonzalez and Herrero \cite{gonzalez2010graphene,gonzalez2009propagating} and observe the presence of degenerate energy levels in the solution of the Dirac equation due to  an external magnetic field. These degenerescences are similar to those  ones arising for Landau levels,  when the graphene quasiparticle is trapped by an external magnetic field. The wormhole is constructed from  a system where one has two graphene-nanotube junctions which play the role of asymptotically flat spaces connected through a carbon nanotube with a zig-zag boundary mimicking a bridge. This wormhole-like geometry is called a wormbridge. In order set up a graphene wormbridge, first of all, it is necessary to make a hole in a  graphene layer with an hexagonal shape. Then, new atoms are connecting with the atoms at  the boundary in vertical direction, so that a zig-zag pattern  is created, and  hence it is possible to connect a graphene layer with a zig-zag nanotube. A direct consequence of this construction is the emergence of six heptagonal rings  alternating with hexagonal rings  in the junction. The construction described by Gonz\'alez and Herrero \cite{gonzalez2010graphene} can be extended for a hole with a large radius, keeping the same number of defects. For convenience, we take the continuum limit for the graphene wormhole considering  the nanotube radius much larger than its length. In this way, we do not need to take into account soft variations in the curvature of the graphene-nanotube junction.

This work is organized as follows. In Section II we present a brief review of the general aspects of the problem. We draw some relevant points about the continuum limit approximation in graphene and on topological defects acting as gauge fields. We also present in this section the geometric description of a graphene wormhole. In Section III, we describe the quantum dynamics of a quasiparticle in the graphene wormhole in the presence of an external magnetic field. The magnetic field is introduced  via minimal coupling. In Section IV, we solve the Weyl equation for the two charts of the graphene wormhole, obtaining the  junction conditions  between them in order to find the correct energy spectrum. Finally, in Section V we present the final discussions  of our results.

\section{Generalities}

\subsection{Topological Defects in Graphene}

In order to describe graphene within a wormhole geometry, where there are effects arising from the curvature of space, we need to understand how the presence of topological defects affect the physical properties of graphene. First of all, it is necessary to discuss its description in flat spaces, and then implement the curved spacetime elements. Graphene is a honeycomb structure  composed of carbon atoms at each of its hexagonal vertices, where each unit cell has two carbon atoms \cite{bena2009remarks}. In fact,  one can consider the honeycomb lattice as a superposition of two $\mathcal{A/B}$ triangular sublattices. Each carbon atom present in the graphene lattice has four valence electrons, where three of these electrons form $\sigma$-bonds and are responsible for the elasticity of the lattice. The remaining valence electron forms $\pi$-bonds responsible for the electronic properties of graphene, which can be obtained through the tight-binding model. The  tight-binding Hamiltonian provides the energy spectra  taking into account that the valence electron can hop to the nearest carbon atoms neighbors, and can be expressed in real space as \cite{gonzalez1993electronic, vozmediano2010gauge}:
\begin{equation}
H = -t\left(\sum_{\langle i,j\rangle} c^{\dagger}_{\mathcal{A}i}\cdot c_{\mathcal{B}j} + h.c.\right),
\label{2.B.1}
\end{equation}
where the creation and annihilation operators $c^{\dagger}_{\mathcal{A}i}$ and $c_{\mathcal{B}j}$, respectively,  describe the probability of a valence electron to hop from $\mathcal{A}$ site to $\mathcal{B}$ nearest-neighbor site (or vice versa) by means of the hopping parameter $t$.

\begin{figure}[t!]
 \subfigure[$~$3D visualization of the dispersion relation of graphene given by equation (\ref{2.B.2}).]{\includegraphics[width=0.45\linewidth]{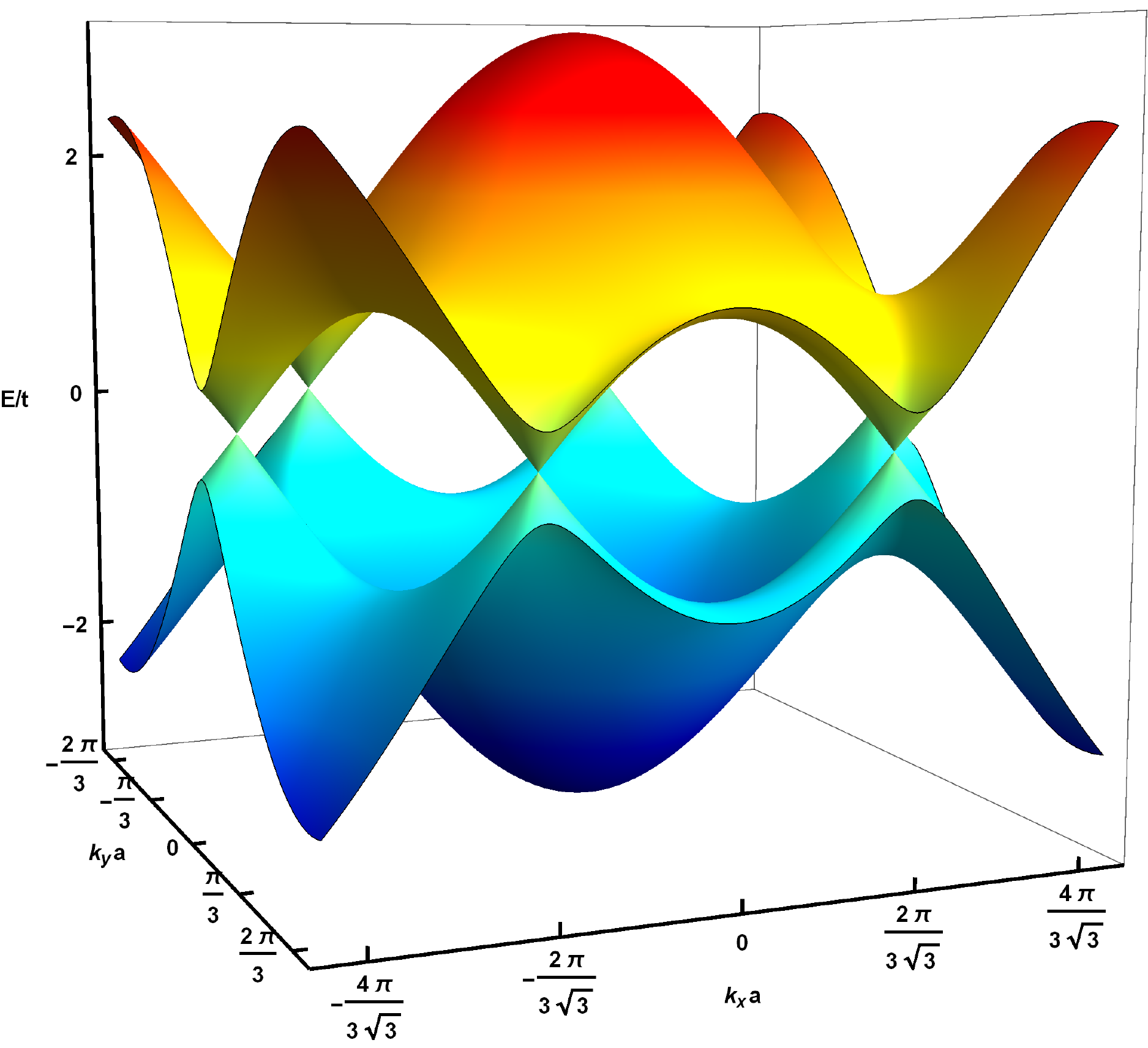}\label{fig1a}} \hspace{5mm}
\subfigure[$~$Hexagonal pattern of a graphene cell. Regions with same color have the same energy and Fermi points $K$ and $K'$.]{\includegraphics[width=0.4\linewidth]{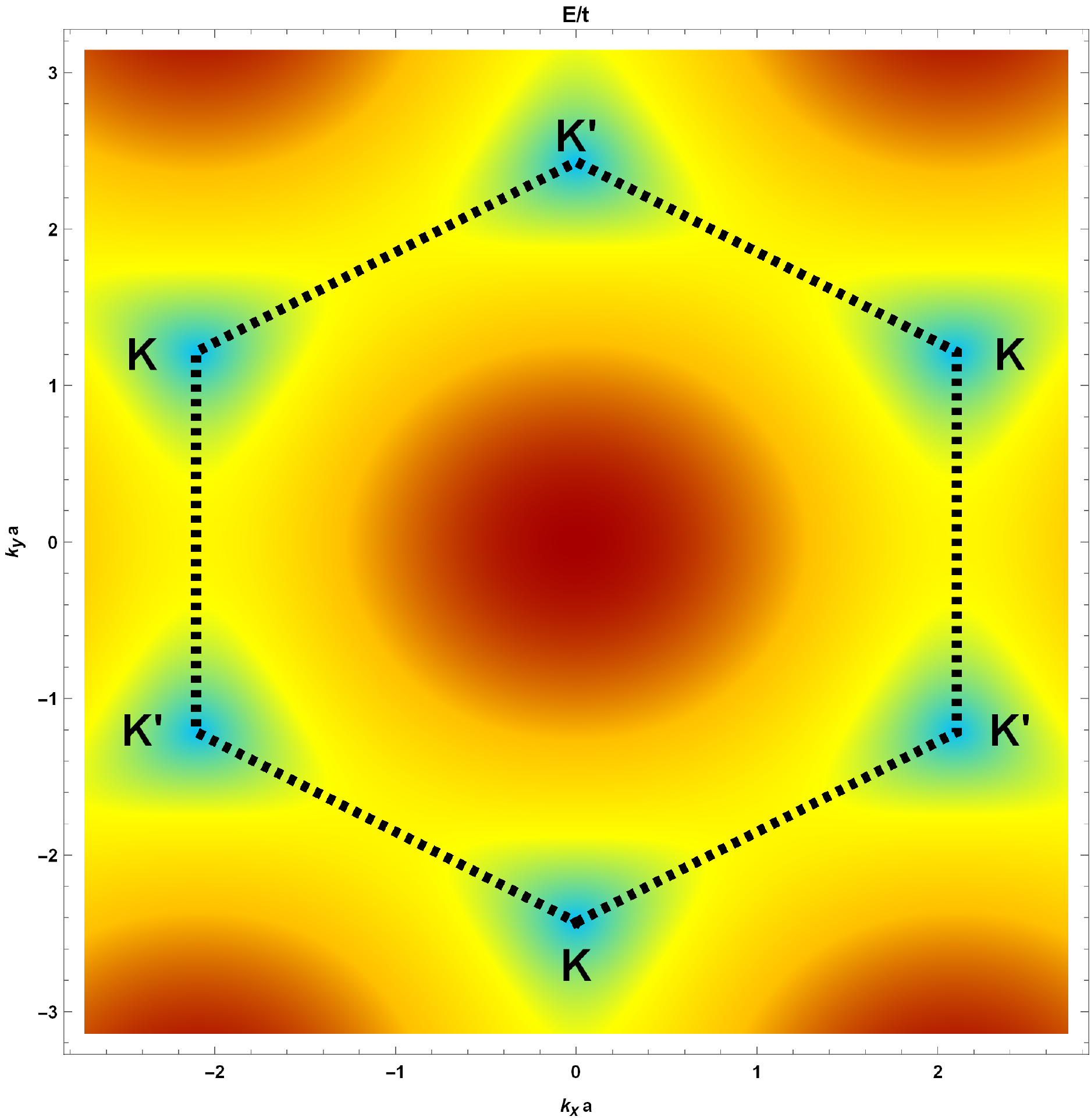}\label{fig1b}} 
  \caption{Band estructure of graphene.\label{fig1} }
  \end{figure}

In order to find the energy band of graphene, one can leave aside the unit cell representation in real space and move to the first Brillouin zone representation in momentum space. After diagonalizing the resulting Hamiltonian, the energy band of graphene becomes
\begin{equation}
E(\vec{k}) = \pm t\sqrt{1+4\cos^2\left(\frac{\sqrt{3}}{2}k_x a\right)+4\cos\left(\frac{\sqrt{3}}{2}k_x a\right)\cos\left(\frac{3}{2}k_y a\right)},
\label{2.B.2}
\end{equation}  
depicted in Fig.\ref{fig1a}, where the parameter $a$ is the interatomic distance between two carbon atoms in the graphene lattice. Since the graphene lattice is described by two atoms in its primitive cell, its reciprocal lattice is described by two moments, denoted by $ K $ and $ K '$, known as Fermi points, so that the first Brillouin zone must be hexagonal, as shown in Fig.\ref{fig1b}. When one considers $E(\vec{k})=0$ in equation (\ref{2.B.2}) the valence band is completely filled, and each corner connects the valence band with conduction band at six points, which are the Fermi points in the hexagon of the reciprocal lattice \cite{bena2009remarks, neto2009electronic}. In the continuum limit, when the electrons are close to the Fermi points (for any of the Fermi points $ K $ or $ K '$), its dispersion relation in the low excitation regime is linear and has a conical shape. Thus, graphene in this regime can be described by the relativistic (2+1)-dimensional massless Dirac equation \cite{bena2009remarks, gonzalez1993electronic}
\begin{equation}
H = \hslash v_f\left(\begin{array}{cc} 0 & k_x - ik_y\\ k_x + ik_y & 0 \end{array}\right) = -i\hslash v_f \vec{\sigma}\cdot\vec{k},
\label{2.B.3}
\end{equation}
where $v_f = \frac{3}{2}ta \approx 10^6\, m/s$ is  the Fermi velocity and $\vec{\sigma}$ is a vector representation for the Pauli matrices. Thus, the interaction between the valence electron and the Fermi points  is described in terms of a quasi-particle obeying the massless Dirac equation such that its eigenvector is a spinor with a pseudospin represented by the sublattices $\mathcal{A/B}$. 

Due to the presence of pentagons and heptagons in the graphene lattice, the atoms  cannot be distributed in two triangular sublattices. These types of defects are known as disclinations. The formation and counting of such defects in a graphene lattice was described by Volterra through "cut and glue" processes \cite{volterra1907equilibre, puntigam1997volterra}. In a graphene lattice one can construct a pentagon defect by cutting a 60 degrees sector and gluing its borders. It is  also possible to construct a heptagon defect by the insertion of a 60 degrees sector in the lattice. The existence of these kinds of defects is responsible for the non-zero curvature in the lattice and, in particular, the pentagon defects generate a positive curvature, whereas the heptagon defects  correspond to a negative curvature. Previously, Katanaev and Volovich proposed a geometric formulation to describe defects in solids \cite{katanaev1992theory,
  katanaev2005geometric}. Several  applications of geometric theory of  defects to quantum dynamics in the presence of topological defect can be found in Refs. \cite{furtado1994landau,furtado1999landau,de2001landau,bakke2009landau,furtado1994binding}. An important point to be drawn in order to show the equivalence between a gravity theory and the theory of defects in solids is that in the continuum description disclinations and dislocations are related to curvature and torsion respectively. It is also worth mentioning that the presence of pentagons and heptagons compels the connection between two sites of the same sublattice, i.e., for example, when the spinor is transported around the defect, at some point it hops from the $\mathcal{A}$ site to another site of the same type $\mathcal{A}$. It implies the mixing of $K$ and $K'$ components of the spinor and, therefore, it is necessary to introduce a $SU(2)$ gauge field, which acts  in a way similar to that one occurring within the Aharonov-Bohm effect to compensate this mixing \cite{pachos2009manifestations}. In polar coordinates, the effective gauge field is given by
\begin{equation}
e^{i\oint_C A_\theta d\theta} = \left(\begin{array}{cc} 0 & 1\\ -1 & 0
\end{array}\right),
\label{2.B.4}
\end{equation}
where the non-Abelian potential has the form
\begin{equation}
A_\theta = \frac{\Phi}{2\pi}\tau_2,
\label{2.B.5}
\end{equation}
$\tau_2$ is one of the Pauli matrices and $\Phi = \pi/2$. In this way, one can observe that the presence of pentagons and heptagons in the graphene lattice can be mimicked through the fictitious gauge flux $ \Phi$ in the topological defect.

\subsection{Graphene Wormhole Geometry}

Now  let us discuss how wormhole geometry is constructed from carbon nanostructures. The graphene wormhole setup is obtained from two graphene sheets linked by carbon nanotube with zig-zag boundary playing the role of a  bridge. To make such  a construction, it is necessary to make a hole with a hexagonal shape on each sheet of graphene, and in this way it is possible to connect the atoms at the hexagonal edges with new atoms in the vertical direction, creating a zig-zag pattern, which provides a way to connect the graphene sheet with a zig-zag nanotube, which is known as graphene-nanotube junction. A direct consequence is the emergence of six heptagons rings at each junction \cite{gonzalez2009propagating}. The geometry of the graphene wormhole arises when we take the continuum limit of this model. In particular, we consider the radius of the hole much larger than the nanotube length, so we discard soft curvature effects, i.e., the curvature is zero throughout the space except at the junction. As usual, these geometries are described by two coordinate charts to cover the upper and lower sheets. Following \cite{gonzalez2010graphene,gonzalez2009propagating}, the charts are chosen in such a way that the polar coordinates can be implemented as follows: the coordinates $(r_{-},\theta_{-})$ and $(r_{+},\theta_{+})$ covers the lower  sheet with metric $g_{ab}^{-}=\text{diag}\left(1, r_{-}^2\right)$, and the upper one -- with metric $g_{ab}^{+}= \text{diag}\left(1, r_{+}^2\right)$ - branches of the geometry, respectively. The radial coordinates $r_{-}$ and $r_{+}$  cover the range $[R,\infty)$, where $R$ represents the ``throat'' of the wormhole. By performing the coordinate transformation
\begin{equation}
r_{-}=R^2/r_{+},~~\text{with}~~r_{+}\geq R,
\label{TR1}
\end{equation}
one can map the upper sheet over the hole defined by the wormhole throat to the lower sheet.  Thus, the complete geometry of the problem can be described in terms of a single variable, denoted now by $r$ (by dropping the minus sign in $r_{-}$), which now runs from $0$ to $\infty$. In other words, it is possible to cover the whole wormhole space with a monotonic variable (one chart) instead of two, as a direct consequence of the metric having  $ \mathbb{Z}_ 2$ - symmetry \cite{visser1995lorentzian}. After all this discussion, the resulting metric becomes
\bes\label{splitmetric}\ben
ds^2&=&dr^2 + r^2 d\theta^2~~~,~~\text{for}~~ r\geq R,\\
ds^2&=&\left(\frac{R}{r}\right)^4\left(dr^2 + r^2 d\theta^2\right),~~\text{for}~~r\leq R.
\een\ees
Note that both sectors of the above expressed geometry coincide at $ r = R $, so the metric is regular everywhere. One can also rewrite the line element of the model in terms of a conformal geometry as follows
\begin{equation}
ds^2 = \Omega^2 (r)\left(dr^2 + r^2 d\theta^2\right),
\label{2.3}
\end{equation} 
where $\Omega(r)$ is a conformal factor defined by
\begin{eqnarray}
\Omega(r) = \left(\frac{R}{r}\right)^2\Theta(R - r) + \Theta(r - R)
\label{2.4}
\end{eqnarray}
and $\Theta(x)$ is the Heaviside step function.

In this work we adopt the Cartan formalism for reasons already indicated in the Introduction  and which also will be better clarified soon. Within this approach, we can define on the $2$-dimensional manifold an  ortonormal co-frame, sometimes called \textit{vielbein} or \textit{zwelbein}, given by  $\hat{\theta}^A = e^{A}_{\ \mu}(x)\ dx^\mu$, where capital Latin letters label flat indices. For the conformal metric presented  in equation \eqref{2.3} one can take an orthonormal co-frame as follows:
\bes\ben
\theta^{r}&=&\Omega(r) dr,\\
\theta^{\theta}&=& d\theta,
\een\ees
where $ds^{2}=\delta_{AB}\theta^{A}\theta^{B}$, with $\delta_{AB}=\text{diag}(+1,+1)$. The next step is to find the spin connections by solving the Cartan structure equations $d\theta^A + \omega^{A}_{\ B}\wedge\theta^B = T^{A}$. By using  the torsionless gauge $T^{A}=0$ the non-vanishing components of the spin connections becomes
\bes\label{spinconnec}\ben
\omega^{\ 1}_{\theta\ 2} &=& -\omega^{\ 2}_{\theta\ 1} = -1,~~\text{for}~~r\geq R,\\
\omega^{\ 1}_{\theta\ 2} &=& -\omega^{\ 2}_{\theta\ 1} = +1,~~\text{for}~~ r\leq R. 
\een\ees
The Cartan approach is the most natural way of coupling fermions to gravity, and from this  viewpoint we look for solutions of the massless  Dirac equation in 2 + 1  dimensions in the background geometry described in this section.

\section{Graphene wormhole trapped by external magnetic field}
As it was pointed out in the previous sections, in the continuum limit it is possible to describe the behavior of graphene quasiparticles through a massless Dirac equation in $(2 + 1)$ dimensions. We are now interested in systems  involving two gauge fields introduced  via minimal coupling: the first one (denoted by $A_{\mu}$) is related to defects generated by the presence of heptagon rings in the graphene lattice, and the second one (denoted by $A_{\mu}^B$) comes from the  imposing of an external magnetic field. The resulting Weyl equation is given by
\begin{eqnarray}
iv_{f}\ \sigma^\mu\left(\nabla_\mu - iA_{\mu} - ie A^{B}_{\mu}\right)\Psi = 0,
\label{3.1}
\end{eqnarray}
where $v_f$ is the Fermi velocity,  $\nabla_\mu = \partial_\mu + \Gamma_\mu$ is the covariant derivative, with $\Gamma_\mu = \frac{1}{8}\omega_{\mu ab}\left[\sigma^a, \sigma^b\right]$ denoting the spinorial connection, and $\lbrace\sigma^i\rbrace$ are  the Pauli matrices.  By using expressions (\ref{spinconnec}$a,b$), a  straightforward inspection shows that the non-vanishing components of the spinorial connection are
\bes\ben
\Gamma_\theta &=& -\frac{i\sigma^3}{2},~~ \text{for} ~r\geq R,\\
\Gamma_\theta &=& +\frac{i\sigma^3}{2},~~\text{for} ~r\leq R.
\een\ees
The gauge field $A_\mu$ arising due to the presence of six heptagon rings at each junction of graphene nanotubes  acts as the fictitious gauge field which, as discussed in the previous section, mimics the topological defects in the graphene lattice. In this way, it is necessary to introduce into the Weyl equation a  term similar to that one present in the expression \eqref{2.B.5}, but now the flux $\Phi$ provides the global contribution of each topological defect in the graphene wormhole lattice. This gauge field is given by
\begin{eqnarray}
A_{\theta} = \pm\frac{\Phi}{2\pi}.
\label{3.3}
\end{eqnarray}
Therefore, equation \eqref{3.3} gives us the Abelian component of the the projection in $\tau^2$ space for the vector potential. The second gauge field $A_\mu^B$ is related to an external magnetic field and expressed as
\bes\ben
A^{B}_{\theta} &=& +\frac{B_0 r^2}{2}, ~~ \text{for} ~r\geq R,\\
 A^{B}_{\theta} &=& -\frac{B_0 R^4}{2r^2}, ~~\text{for} ~r\leq R.
\een\ees

By using the gauge fields described so far we can explicitly write down the Weyl equation \eqref{3.1} for the wave functions inside and outside the wormhole throat to describe the quantum dynamics of graphene wormhole quasiparticle. They are given by
\bes\label{syst1}\ben
\frac{\sigma^0}{v_f}\frac{\partial\Psi}{\partial t} + \sigma^1\left(\frac{\partial}{\partial r}+\frac{1}{2r}\right)\Psi + \frac{\sigma^2}{r}\left(\frac{\partial}{\partial\theta}- iA_\theta - i\frac{eB_0 r^2}{2}\right)\Psi &=& 0,~~\text{for}~~r\geq R,\\[4pt]
\frac{\sigma^0}{v_f}\frac{\partial\Psi}{\partial t} + \frac{\sigma^1 r^2}{R^2}\left(\frac{\partial}{\partial r}-\frac{1}{2r}\right)\Psi + \frac{\sigma^2 r}{R^2}\left(\frac{\partial}{\partial\theta}- iA_\theta + i\frac{eB_0 R^4}{2r^2}\right)\Psi &=& 0, ~~\text{for}~~r\leq R.
\een\ees
The first step to solve the above system is to  choose the following ansatz for the wave function: 
\begin{eqnarray}
\Psi = e^{-i\frac{Et}{\hslash}}\left(\begin{array}{c}\Psi^{\pm}_{A}\\ \Psi^{\pm}_{B}\end{array}\right).
\end{eqnarray}
In this way, the system (\ref{syst1}$a,b$) can be rewritten as
\bes\label{3.7}\ben
 -iv_{f}\hslash\left(\begin{array}{cc}0 & \partial_r +\frac{1}{2r}-\frac{i}{r}\partial_{\theta}\mp\frac{\bar{\Phi}}{2\pi r} - \frac{eB_0 r}{2}\\ \partial_r +\frac{1}{2r}+\frac{i}{r}\partial_{\theta}\pm\frac{\bar{\Phi}}{2\pi r} + \frac{eB_0 r}{2} & 0
\end{array}\right)\left(\begin{array}{c}\psi^{\pm}_{A}\\ \psi^{\pm}_{B}\end{array}\right)&=& E\left(\begin{array}{c}\psi^{\pm}_{A}\\ 
\psi^{\pm}_{B}\end{array}\right),~~ \text{for} ~r\geq R,\qquad\quad\\[4pt]
-iv_{f}\hslash\left(\frac{r}{R}\right)^2\left(\begin{array}{cc}0 & \partial_r -\frac{1}{2r}-\frac{i}{r}\partial_{\theta}\mp\frac{\tilde{\Phi}}{2\pi r} + \frac{eB_0 r}{2}\\ \partial_r -\frac{1}{2r}+\frac{i}{r}\partial_{\theta}\pm\frac{\tilde{\Phi}}{2\pi r} - \frac{eB_0 r}{2} & 0
\end{array}\right)\left(\begin{array}{c}\psi^{\pm}_{A}\\ \psi^{\pm}_{B}\end{array}\right) &=& E\left(\begin{array}{c}\psi^{\pm}_{A}\\ \psi^{\pm}_{B}\end{array}\right).  ~~\text{for} ~r\leq R.\qquad\quad
\een\ees
Note that in the system above we  suggest different effective  fluxes $\bar{\Phi}$ and $\tilde{\Phi}$ for the regions inside and outside the wormhole throat, respectively, since the electron must be affected in different ways by these flows in those regions. In \cite{gonzalez2010graphene}  it is argued that one must have $\bar{\Phi}=-\tilde{\Phi}$. We  note that this relation arises as one of the junction conditions to be imposed on the wave function in order to obtain well-behaved solutions over the entire background geometry.

\subsection{Solving the massless Dirac equation}
We now  present an approach to solve the Weyl equations in the background geometry of the graphene wormhole discussed in the previous section. We split this procedure into two parts referring to the two distinct regions of the  space.

\subsubsection{Outside of  the throat ($r\geq R$)}

First, we solve the Weyl  equation (\ref{3.7}$a$) describing the graphene in the region outside the throat of the wormhole. For this case, it is convenient to take a complementary ansatz  for the wave function, given by
\begin{eqnarray}
\left(\begin{array}{c}\Psi^{\pm}_{A}\\ \Psi^{\pm}_{B}\end{array}\right)_{\left( r\geq R\right)} = e^{ij\theta} \left(\begin{array}{c}f_1(r)\\ f_2(r)\end{array}\right),
\label{4.1.1}
\end{eqnarray}
 thus, we have a pair of  coupled first order differential equations,
\bes\label{4.1.2}\ben
-iv_{f}\hslash\left(\frac{\partial}{\partial r}+\frac{1}{2r}+\frac{j}{r}\mp\frac{\bar{\Phi}}{2\pi r}-\frac{eB_{0}r}{2}\right)f_2(r) &=& E f_1(r);\\
-iv_{f}\hslash\left(\frac{\partial}{\partial r}+\frac{1}{2r}-\frac{j}{r}\pm\frac{\bar{\Phi}}{2\pi r}+\frac{eB_{0}r}{2}\right)f_1(r) &=& E f_2(r).
\een\ees
By decoupling the above system, we find another pair of differential equations, but now they are second order ones, being given in the following compact form: 
\ben
\frac{d^{2}f_{i}(r)}{dr^2} + \frac{1}{r}\frac{df_i(r)}{dr} - \left[\frac{e^2 B_{0}^{2}}{4} r^{2} + \frac{\zeta_{i}^{2}}{r^{2}} - \beta_i\right]f_i(r) = 0,
\label{4.1.3}
\een
where the $i = 1, 2$, and we have introduced the following parameters: 
\bes\label{4.1.5}\ben
\zeta_{1}^{2}=\left(j - \frac{1}{2} \mp \frac{\bar{\Phi}}{2\pi}\right)^2~~&,&~~ \zeta_{2}^{2}=\left(j + \frac{1}{2} \mp \frac{\bar{\Phi}}{2\pi}\right)^2,\\[4pt]
\beta_1 = \frac{E^2}{v_{f}^{2}\hslash^2}\mp \frac{e B_{0}\bar{\Phi}}{2\pi}+jeB_{0} + \frac{eB_{0}}{2}~~&,&~~\beta_2 = \frac{E^2}{v_{f}^{2}\hslash^2}\mp \frac{e B_{0}\bar{\Phi}}{2\pi}+jeB_{0}- \frac{eB_{0}}{2}.
\label{4.1.4}
\een\ees
We can  rewrite equation \eqref{4.1.3} in a more convenient way if we perform the change of variable $\xi = \frac{eB_{0}}{2}r^{2}$, so one finds
\ben
\xi\frac{d^{2}f_i(\xi)}{d\xi^{2}} + \frac{df_i(\xi)}{d\xi} + \left[\frac{\beta_i}{2eB_{0}} - \frac{\xi}{4} - \frac{\zeta_{i}^{2}}{4\xi}\right]f_i(\xi) = 0.
\label{4.1.6}
\een
Equation \eqref{4.1.6} has  the asymptotic solution given by $f(\xi) = e^{-\frac{\xi}{2}}\xi^{\frac{\vert\zeta\vert}{2}} F(\xi)$, where $F(\xi) = {}_{1}F_{1}\left(-\left(\frac{\beta}{2eB_{0}}-\frac{1}{2}-\frac{|\zeta|}{2}\right),\vert\zeta\vert + 1;\xi\right)$  is a confluent hypergeometric function \cite{abramowitz1965handbook}. The hypergeometric confluent equations is 
\begin{eqnarray}
\xi\frac{\partial^{2} F}{\partial \xi^{2}}+\left(\vert\zeta\vert+1-\xi\right)\frac{\partial F}{\partial\xi} + \left(\frac{\beta}{2eB_{0}}-\frac{1}{2}-\frac{|\zeta|}{2}\right) F = 0.
\label{3.1.7}
\end{eqnarray}
Note that here we have a dependence on $|\zeta|$, so we must be careful in dealing with values of $ j $. Thus, the wave function solution (Weyl spinor) found in the region $r \geq R$ is
\begin{eqnarray}
\Psi_{\left(r\geq R,\zeta\geq0\right)} = A_{n,j}\ e^{-i\left(\frac{Et}{\hslash}-j\theta\right)}\ e^{-\frac{\xi}{2}}\left(\begin{array}{c} \xi^{\frac{1}{2}\left(j-\frac{1}{2}\mp\frac{\bar{\Phi}}{2\pi}\right)}\ F\left(-n, j+\frac{1}{2}\mp\frac{\bar{\Phi}}{2\pi}; \xi\right)\\
\xi^{\frac{1}{2}\left(j+\frac{1}{2}\mp\frac{\bar{\Phi}}{2\pi}\right)}\ F\left(-n, j+\frac{3}{2}\mp\frac{\bar{\Phi}}{2\pi}; \xi\right)
\end{array}\right)
\label{3.1.8}
\end{eqnarray}
for $\zeta \geq 0$ and
\begin{eqnarray}
\Psi_{\left(r\geq R,\zeta\leq0\right)} = B_{n,j}\ e^{-i\left(\frac{Et}{\hslash}-j\theta\right)}\ e^{-\frac{\xi}{2}}\left(\begin{array}{c} \xi^{\frac{1}{2}\left(-j+\frac{1}{2}\pm\frac{\bar{\Phi}}{2\pi}\right)}\ F\left(-n, -j+\frac{3}{2}\pm\frac{\bar{\Phi}}{2\pi}; \xi\right)\\
\xi^{\frac{1}{2}\left(-j-\frac{1}{2}\pm\frac{\bar{\Phi}}{2\pi}\right)}\ F\left(-n, -j+\frac{1}{2}\pm\frac{\bar{\Phi}}{2\pi}; \xi\right)
\end{array}\right)
\label{3.1.8a}
\end{eqnarray}
for $\zeta \leq 0$, where $A_{n,j}$ and $B_{n,j}$  are normalization constants.

\subsubsection{Inside of the throat ($r\leq R$)}

 As a continuation, we now solve the Weyl equation (\ref{3.7}$b$),  describing the behavior of graphene inside of the wormhole throat.  Similarly to the previous subsection, we start from a specific choice for the wave function to find a system of first order differential equations which can be decoupled in two second order equations. By using  the ansatz
\begin{eqnarray}
\left(\begin{array}{c}\Psi^{\pm}_{A}\\ \Psi^{\pm}_{B}\end{array}\right)_{\left( r\leq R\right)} = e^{i\tilde{j}\theta} \left(\begin{array}{c}g_1(r)\\ g_2(r)\end{array}\right),
\label{4.2.1}
\end{eqnarray}
one finds the pair of coupled first order equations given by
\bes\ben
-iv_{f}\hslash\left(\frac{r}{R}\right)^{2}\left(\frac{\partial}{\partial r}-\frac{1}{2r}+\frac{\tilde{j}}{r}\mp\frac{\tilde{\Phi}}{2\pi r}+\frac{eB_{0}R^{4}}{2r^{3}}\right)g_{2}(r) &=& E g_{1}(r),\\
-iv_{f}\hslash\left(\frac{r}{R}\right)^{2}\left(\frac{\partial}{\partial r}-\frac{1}{2r}-\frac{\tilde{j}}{r}\pm\frac{\tilde{\Phi}}{2\pi r}-\frac{eB_{0}R^{4}}{2r^{3}}\right)g_{1}(r) &=&E g_{2}(r).
\een\ees
Now, before decoupling the above system, we need to perform a change of variables in order to find equations  allowing to obtain hypergeometric functions. By using $\tilde{r} = \frac{R^2}{r}$ we find
\bes\label{4.2.2}\ben
i\left(\frac{\partial}{\partial \tilde{r}}+\frac{1}{2\tilde{r}}-\frac{\tilde{j}}{\tilde{r}}\pm\frac{\tilde{\Phi}}{2\pi \tilde{r}}-\frac{eB_{0}\tilde{r}}{2}\right)g_{2}(\tilde{r}) &=& \left(\frac{E}{v_{f}\hslash}\right) g_{1}(\tilde{r}),\\
i\left(\frac{\partial}{\partial\tilde{r}}+\frac{1}{2\tilde{r}}+\frac{\tilde{j}}{\tilde{r}}\mp\frac{\tilde{\Phi}}{2\pi\tilde{r}}+\frac{eB_{0}\tilde{r}}{2}\right)g_{1}(\tilde{r}) &=& \left(\frac{E}{v_{f}\hslash}\right) g_{2}(\tilde{r}).
\een\ees
After decoupling this system and making some manipulations, we have
\ben
\frac{d^{2}g_{i}(\tilde{r})}{d\tilde{r}^2} + \frac{1}{\tilde{r}}\frac{dg_i(\tilde{r})}{d\tilde{r}} - \left[\frac{e^2 B_{0}^{2}}{4}\tilde{r}^2 + \frac{\zeta_{i}^{2}}{\tilde{r}^{2}} - \beta_i\right]g_i(\tilde{r}) = 0,
\label{4.2.3}
\een
where, again, $i= 1, 2$ and
\bes\label{4.2.5}\ben
\tilde{\zeta}_{1}^{2}=\left(\tilde{j} + \frac{1}{2} \mp \frac{\tilde{\Phi}}{2\pi}\right)^2~~&,&~~ \tilde{\zeta}_{2}^{2}=\left(\tilde{j} - \frac{1}{2} \mp \frac{\tilde{\Phi}}{2\pi}\right)^2,\\[4pt]
\tilde{\beta}_1 = \frac{E^2}{v_{f}^{2}\hslash^2}\pm \frac{e B_{0}\tilde{\Phi}}{2\pi}-\tilde{j} eB_{0} + \frac{eB_{0}}{2}~~&,&~~\tilde{\beta_2} = \frac{E^2}{v_{f}^{2}\hslash^2}\pm \frac{e B_{0}\tilde{\Phi}}{2\pi}-\tilde{j}eB_{0}- \frac{eB_{0}}{2}.
\label{4.2.4}
\een\ees
By performing a second coordinate change in the system \eqref{4.2.3}, now given by $\chi = \frac{eB_{0}}{2}\tilde{r}^2$, one finds
\begin{equation}
\chi\frac{\partial^{2} g_{i}(\chi)}{\partial \chi^{2}}+\frac{\partial g_{i}(\chi)}{\partial\chi} + \left(\frac{\tilde{\beta}_i}{2eB_{0}}-\frac{\chi}{4}-\frac{\tilde{\zeta}_{i}^2}{4\chi}\right)g_{i}(\chi)= 0,
\label{4.2.6}
\end{equation}
and this equation has asymptotic solution given by $g_{i}(\chi) = e^{-\frac{\chi}{2}}\chi^{\frac{\vert\tilde{\zeta}\vert}{2}} \tilde{F}(\chi)$ with $\tilde{F}(\chi) = {}_1F_{1}\left(-\left(\frac{\tilde{\beta}}{2eB_{0}}-\frac{1}{2}-\frac{\tilde{|\zeta|}}{2}\right),\vert\tilde{\zeta}\vert + 1;\chi\right)$  corresponding to the confluent hypergeometric function \cite{abramowitz1965handbook}, which is a solution of the following confluent hypergeometric differential equation: 
\begin{eqnarray}
\chi\frac{\partial^{2}\tilde{F}}{\partial \chi^{2}}+\left(\vert\tilde{\zeta}\vert+1-\chi\right)\frac{\partial \tilde{F}}{\partial\chi} + \left(\frac{\tilde{\beta}}{2eB_{0}}-\frac{1}{2}-\frac{\tilde{|\zeta|}}{2}\right)\tilde{F} = 0.
\label{3.2.7}
\end{eqnarray}
Thus, the general solution for the Weyl spinor,  with $r \leq R$, is given by:
\begin{eqnarray}
\Psi_{\left(r\leq R,\tilde{\zeta}\geq0\right)}= \tilde{A}_{\tilde{n},\tilde{j}}\ e^{-i\left(\frac{Et}{\hslash}-\tilde{j}\theta\right)}\ e^{-\frac{\chi}{2}}\left(\begin{array}{c} \chi^{\frac{1}{2}\left(\tilde{j}+\frac{1}{2}\mp\frac{\tilde{\Phi}}{2\pi}\right)}\ \tilde{F}\left(-\tilde{n}, \tilde{j}+\frac{3}{2}\mp\frac{\tilde{\Phi}}{2\pi}; \chi\right)\\
\chi^{\frac{1}{2}\left(\tilde{j}-\frac{1}{2}\mp\frac{\tilde{\Phi}}{2\pi}\right)}\ \tilde{F}\left(-\tilde{n}, \tilde{j}+\frac{1}{2}\mp\frac{\tilde{\Phi}}{2\pi}; \chi\right)
\end{array}\right),
\label{3.1.8b}
\end{eqnarray}
for $\tilde{\zeta}\geq 0$ and
\begin{eqnarray}
\Psi_{\left(r\leq R,\tilde{\zeta}\leq0\right)} = \tilde{B}_{\tilde{n},\tilde{j}}\ e^{-i\left(\frac{Et}{\hslash}-\tilde{j}\theta\right)}\ e^{-\frac{\chi}{2}}\left(\begin{array}{c} \chi^{\frac{1}{2}\left(-\tilde{j}-\frac{1}{2}\pm\frac{\tilde{\Phi}}{2\pi}\right)}\ \tilde{F}\left(-\tilde{n}, -\tilde{j}+\frac{1}{2}\pm\frac{\tilde{\Phi}}{2\pi}; \chi\right)\\
\chi^{\frac{1}{2}\left(-\tilde{j}+\frac{1}{2}\pm\frac{\tilde{\Phi}}{2\pi}\right)}\ \tilde{F}\left(-\tilde{n}, -\tilde{j}+\frac{3}{2}\pm\frac{\tilde{\Phi}}{2\pi}; \chi\right)
\end{array}\right),
\label{3.1.8c}
\end{eqnarray}
for $\tilde{\zeta}\leq 0$,  and $\tilde{A}_{\tilde{n},\tilde{j}}$ and $\tilde{B}_{\tilde{n},\tilde{j}}$ are normalization constants.

\subsection{Junction conditions, energy levels and complete wave function solution}

In the previous sections we found the solutions for electrons in the inner and outer regions of the wormhole throat. Now it is necessary to check  which junction conditions are required for consistency of these solutions. The junction conditions we use here for the Weyl spinors  inside and  outside of the wormhole throat are given by imposing that $n=\tilde{n}$ (i.e., the wave functions  correspond to the same quantum numbers  for both these cases) and
\bes\label{4.3.1}\ben
\left.\Psi_{\left(r\geq R,\zeta\geq0\right)}\right|_{r=R}&=&\left.\Psi_{\left(r\leq R,\tilde{\zeta}\leq0\right)}\right|_{r=R};\\
\left.\Psi_{\left(r\geq R,\zeta\leq0\right)}\right|_{r=R}&=&\left.\Psi_{\left(r\leq R,\tilde{\zeta}\geq0\right)}\right|_{r=R}.
\een\ees
By using the above constraints, we can find the other equivalences between parameters  used in the solutions, given by
\begin{equation}
j=-\tilde{j},~\bar{\Phi}=-\tilde{\Phi},~\tilde{B}_{n,-j}=e^{2ij\theta}A_{n,j},~\text{and}~\tilde{A}_{n,-j}=e^{2ij\theta}B_{n,j}.
\label{4.3.2}
\end{equation}
These identifications and the requirement that the confluent hypergeometric  function truncates, i. e., it becomes a $n$ degree polynomial, implies that the energy levels of the Weyl equation \eqref{3.1}  become
\begin{equation}\label{energylevels}
E_{n,j,\epsilon}=\pm v_f\hslash\sqrt{2eB_0\left[n+\frac{1}{2}\left|j-\frac{1}{2}-\epsilon\frac{\bar{\Phi}}{2\pi}\right|-\frac{1}{2}\left(j-\frac{1}{2}-\epsilon\frac{\bar{\Phi}}{2\pi}\right)\right]},
\end{equation}
\begin{figure}[ht!]
 \subfigure[$~$Landau Levels for $\bar{\Phi}=\pi$]{\includegraphics[width=0.48\linewidth]{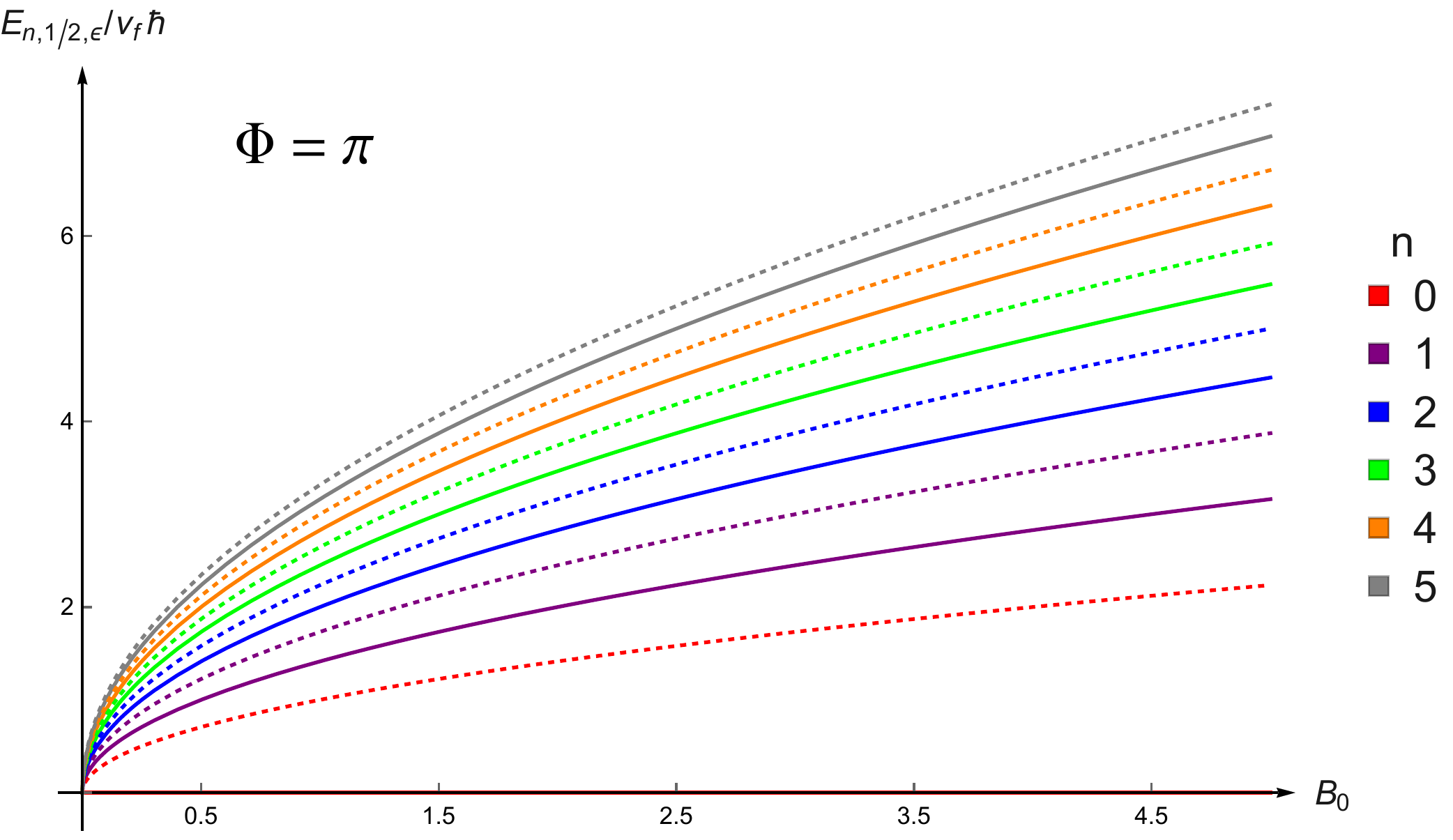}\label{fig2a}}\hspace{5mm}
\subfigure[$~$Landau Levels for $\bar{\Phi}=3\pi$.]{ \includegraphics[width=0.48\linewidth]{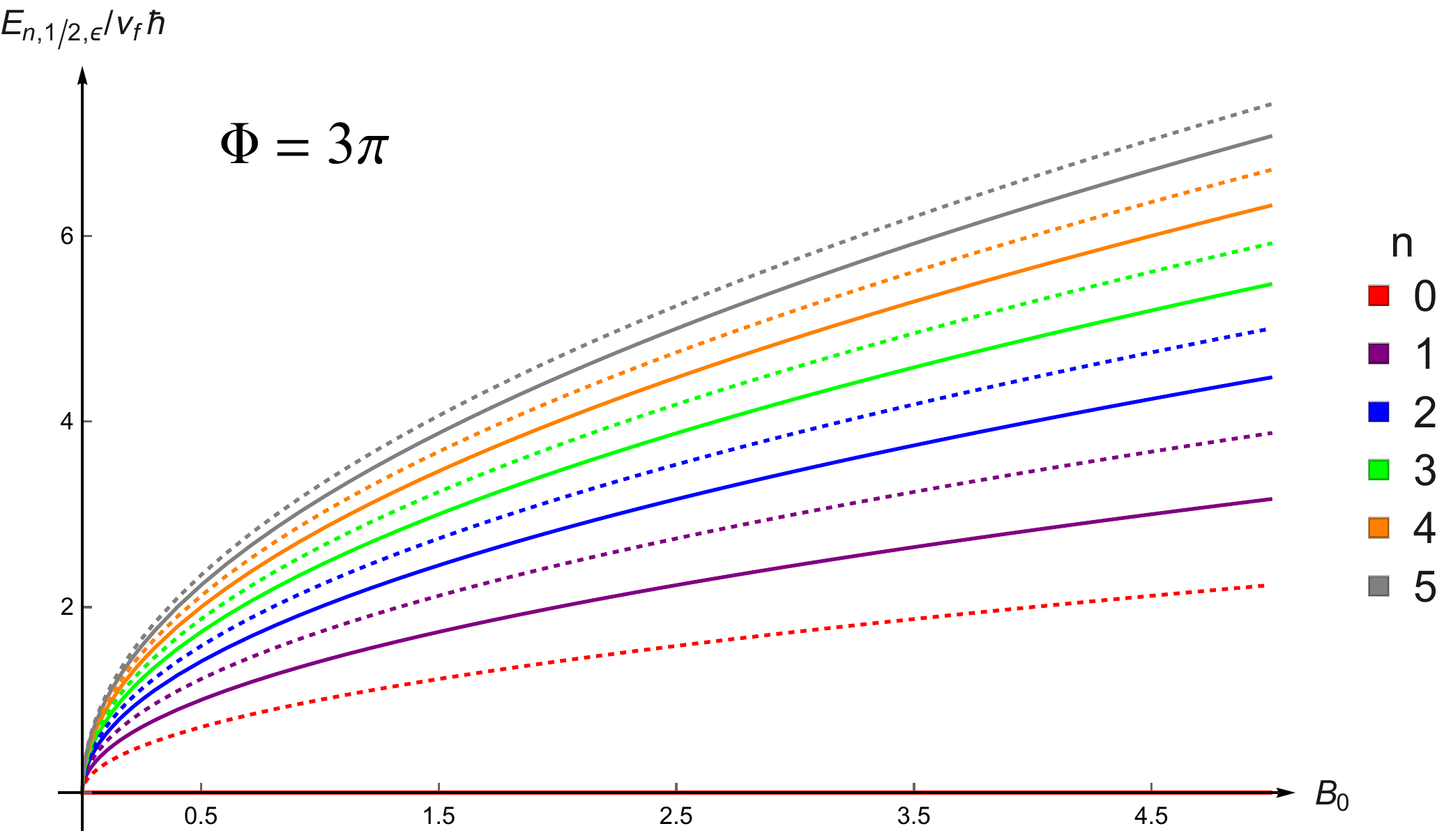}\label{fig2b}} 
  \caption{Evolution of the Landau levels \eqref{energylevels} as a function of the homogeneous magnetic field $B_0$. The dotted lines are for cases where $\epsilon=1$ and solid lines are for $\epsilon=-1$.\label{fig2} }
  \end{figure}
where $n = 0, 1, 2...$ is  integer, $j$ is a half-integer and $\epsilon=\pm1$. The energy levels allowed by the equation \eqref{energylevels} show that the presence of an external magnetic field is essential for the existence of bound states exhibiting  effects  analogous to relativistic Landau levels \cite{bakke2009relativistic} in the graphene wormhole with a relation of type  $ E_n \propto \sqrt{B_0} $ depicted in FIG.\ref{fig2}($a,b$). In this analogy, the $\mathcal{A}/\mathcal{B}$ sublattices play the role of pseudospin, so that the presence of topological defects breaks the degeneracy in doublets. Thus, the behavior of the energy  spectrum \eqref{energylevels} as a function of the external magnetic field is similar to relativistic  Landau levels, but with the difference that  the the breaking of the degeneracy in doublets occurs now. Note that there is no $R$-dependence on energy levels, which is due to the fact that we are dealing with massless fermions in a conformal geometry. The global effect due to the topological defects given by $\bar{\Phi}$ flux can emerge in two ways when one  suggests that the topological defects are regularly spreaded out over the graphene-nanotube junction \cite{lammert2004graphene}. When the distance between two heptagon rings is a multiple of 3, the overall effect is the sum of the individual fluxes given by equation \eqref{2.B.5} and then the total flux is $\bar{\Phi}=3\pi$. For the case where the distance between two heptagon rings is not multiple of 3, each pair of heptagon rings gives a contribution of $\pi/3$ and hence the total flux is $\bar{\Phi }= \pi$. In FIG.\ref{fig3}($a,b$) we show the energy spectrum of \eqref{energylevels} for fixed  $B_0$ and $j$ and different values of $n, \bar{\Phi}$ and $\epsilon$.For $j= -\frac{1}{2}$ and $\epsilon = -1$  one can see the degeneracy breaking for energy levels in doublets for $\bar{\Phi}=3\pi$ and $\bar{\Phi}=\pi$. In this case we do not observe degeneracy for different values $n$. This effect is due the presence of topological defects and the ground state is observed $n=0$ and $\bar{\Phi}=3\pi$. For $j = -\frac{1}{2}$ and $\epsilon = 1$ there is also a degeneracy breaking in doublets similar to the previous case, but with the difference that the lowest energy level now occurs in the case where the  fictitious flux is given by $\bar{\Phi}=\pi$. In the figure FIG.\ref{fig3b} there is a degeneracy between consecutive $ n$ values, where the energy level of highest value for some $ n $ ($\bar{\Phi}=3\pi$)  is equal to the energy level of lowest value for $ n + 1 $ ($\bar{\Phi}=\pi$). These emergent effects arise  because of the way we organize the heptagones arising in the presence of negative disclinations in the construction of graphene wormhole. For both cases, there is zero mode only for a positive total flux. Finally, the complete wave function solutions becomes
\bes\ben
\Psi_{\left(r\geq R,\zeta\geq0\right)} &=& A_{n,j}\ e^{-i\left(\frac{Et}{\hslash}-j\theta\right)}\ e^{-\frac{\xi}{2}}\left(\begin{array}{c} \xi^{\frac{1}{2}\left(j-\frac{1}{2}\mp\frac{\bar{\Phi}}{2\pi}\right)}\ F\left(-n, j+\frac{1}{2}\mp\frac{\bar{\Phi}}{2\pi}; \xi\right)\\[4pt]
\xi^{\frac{1}{2}\left(j+\frac{1}{2}\mp\frac{\bar{\Phi}}{2\pi}\right)}\ F\left(-n, j+\frac{3}{2}\mp\frac{\bar{\Phi}}{2\pi}; \xi\right)
\end{array}\right);\\
\Psi_{\left(r\leq R,\tilde{\zeta}\leq0\right)} &=& A_{n,j}\ e^{-i\left(\frac{Et}{\hslash}-j\theta\right)}\ e^{-\frac{\chi}{2}}\left(\begin{array}{c} \chi^{\frac{1}{2}\left(j-\frac{1}{2}\pm\frac{\tilde{\Phi}}{2\pi}\right)}\ F\left(-n, j+\frac{1}{2}\pm\frac{\tilde{\Phi}}{2\pi}; \chi\right)\\
\chi^{\frac{1}{2}\left(j+\frac{1}{2}\pm\frac{\tilde{\Phi}}{2\pi}\right)}\ F\left(-n, j+\frac{3}{2}\pm\frac{\tilde{\Phi}}{2\pi}; \chi\right)
\end{array}\right),
\een\ees
and 
\bes\ben
\Psi_{\left(r\geq R,\zeta\leq0\right)} &=& B_{n,j}\ e^{-i\left(\frac{Et}{\hslash}-j\theta\right)}\ e^{-\frac{\xi}{2}}\left(\begin{array}{c} \xi^{\frac{1}{2}\left(-j+\frac{1}{2}\pm\frac{\bar{\Phi}}{2\pi}\right)}\ F\left(-n, -j+\frac{3}{2}\pm\frac{\bar{\Phi}}{2\pi}; \xi\right)\\
\xi^{\frac{1}{2}\left(-j-\frac{1}{2}\pm\frac{\bar{\Phi}}{2\pi}\right)}\ F\left(-n, -j+\frac{1}{2}\pm\frac{\bar{\Phi}}{2\pi}; \xi\right)
\end{array}\right);\\[4pt]
\Psi_{\left(r\leq R,\tilde{\zeta}\geq0\right)}&=& B_{n,j}\ e^{-i\left(\frac{Et}{\hslash}-j\theta\right)}\ e^{-\frac{\chi}{2}}\left(\begin{array}{c} \chi^{\frac{1}{2}\left(-j+\frac{1}{2}\mp\frac{\tilde{\Phi}}{2\pi}\right)}\ F\left(-n, -j+\frac{3}{2}\mp\frac{\tilde{\Phi}}{2\pi}; \chi\right)\\
\chi^{\frac{1}{2}\left(-j-\frac{1}{2}\mp\frac{\tilde{\Phi}}{2\pi}\right)}\ F\left(-n, -j+\frac{1}{2}\mp\frac{\tilde{\Phi}}{2\pi}; \chi\right)
\end{array}\right).
\een\ees
which covers all graphene wormhole surface.

\begin{figure}[t!]
 \subfigure[$~$$\epsilon=-1$]{\includegraphics[width=0.4\linewidth]{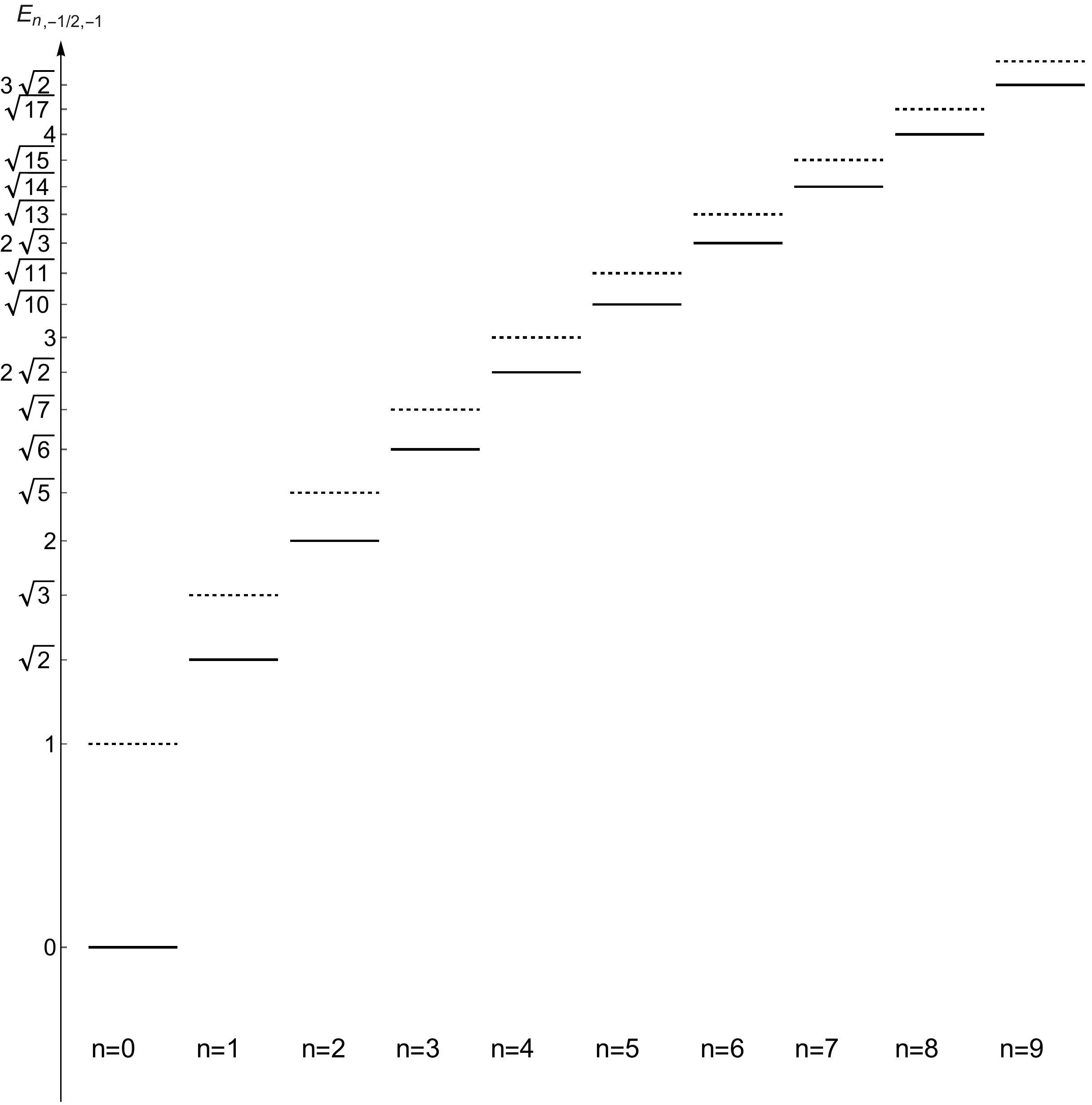}\label{fig3a}} \hspace{10mm}
 \subfigure[$~$$\epsilon=1$]{\includegraphics[width=0.4\linewidth]{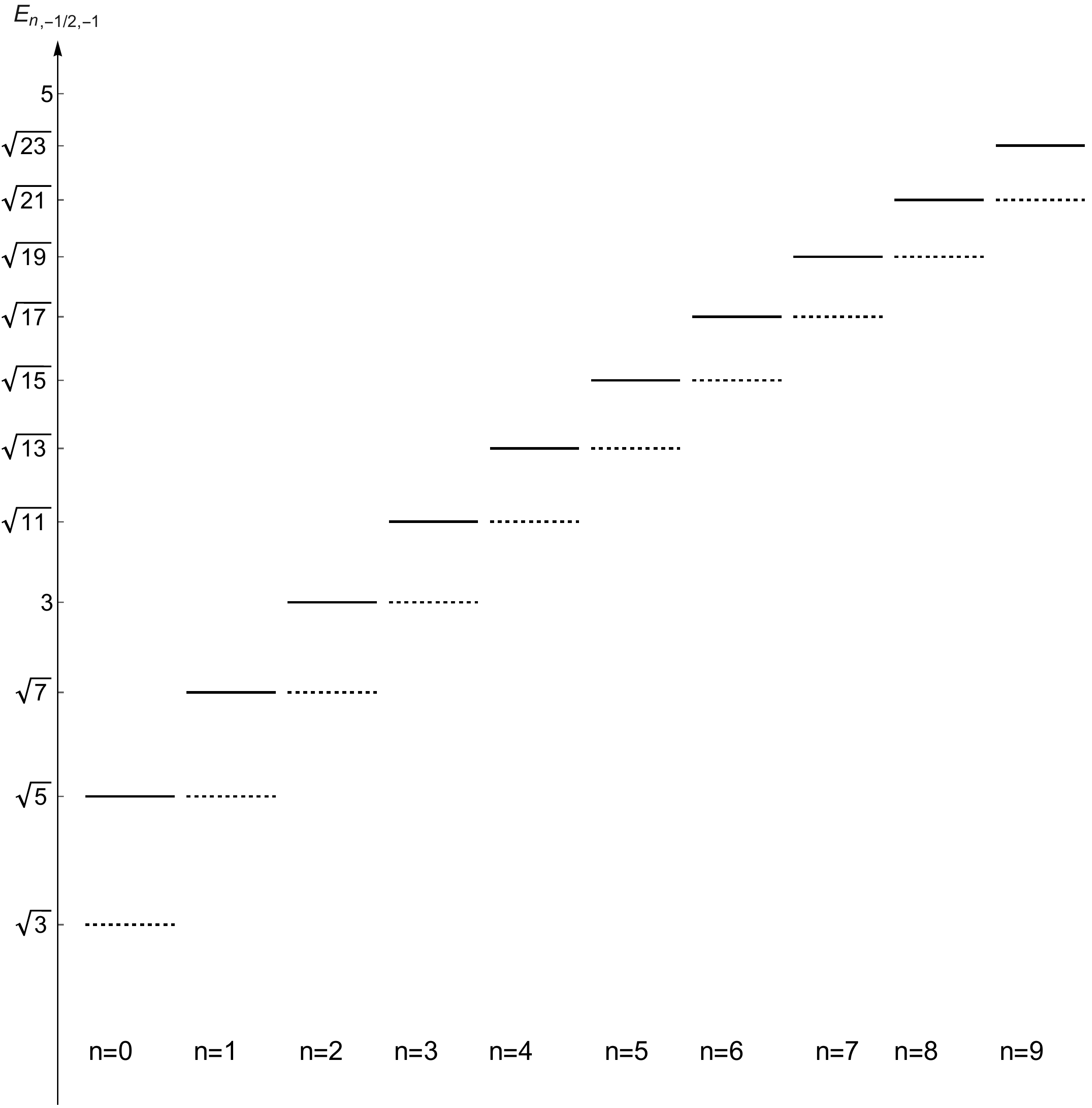}\label{fig3b}} 
  \caption{Energy spectrum \eqref{energylevels} for some values of $n$ for a fixed homogeneous magnetic field $B_0$  and different values of $\epsilon$. The dotted lines are for cases where $\bar{\Phi}=\pi$ and solid lines are for $\bar{\Phi}=3\pi$.\label{fig3} }
  \end{figure}

\section{Ending comments}
In this work we investigated the quantum dynamics of massless relativistic fermions in the geometric model of  the graphene wormhole proposed in \cite{gonzalez2010graphene} in the presence of an external magnetic field. We solved the Weyl equation for the wormhole metric describing connected graphene sheets in the  continuum limit and found analytical expressions to its wave functions and eigenvalues. The curvature effects of the background geometry are related to the presence of disclinations in the graphene lattice, so we observed its influence on the energy levels along with the external magnetic field. The energy levels of this model are similar to relativistic Landau levels in the presence of topological defects \cite{bueno2012landau,bakke2010relativistic} and evolve as a function of the external magnetic field by a relation of type $ E_n \propto \sqrt{B_0} $ for any value of the orbital $ j $. The topological defects contribution denoted by the fictitious gauge flux $\bar{\Phi}$ also affects the behavior of energy levels, as can be noted in equation \eqref{energylevels} and in the figure FIG.\ref{fig3}($a,b$), acting along with the orbital numbers $j$ in the breaking of degeneracy and splitting the energy levels in doublets. A possible and natural extension of this work may be the study of the Quantum Hall Effect in  a graphene wormhole. This work constitutes an additional effort to understand the behavior of graphene in curved surfaces and the role of topological defects and other auxiliary gauge fields in this context.

\acknowledgments
G. Q. Garcia would like to thank the Brazilian development agencies CNPq and Fapesq-PB for the financial support (Fapesq-PB/CNPq grant, process 300354/2018-5), P. J. Porf\'irio would like to thank the Brazilian development agency CAPES for the financial support (PDE/CAPES grant, process 88881.171759/2018-01), D. C. Moreira would like to thank the Brazilian development agency CNPq for the financial support (PDJ/CNPq grant, process 405747/2017-9), and C. Furtado would like to thank Fapesq-PB, CNPq and CAPES. 

\bibliographystyle{unsrt}
\bibliography{biblio} 

\end{document}